\documentclass[11pt, letterpaper]{article}
\usepackage[utf8]{inputenc}
\usepackage[margin=1.5cm]{geometry}
\usepackage{titlesec}
\usepackage{tabu}
\usepackage{longtable}[=v4.13]%
\usepackage{tabu}%

\usepackage{enumitem}
\usepackage{amssymb}
\usepackage{xcolor}
\usepackage{amsmath}
\usepackage{verbatim}
\usepackage[]{graphicx}
\newlist{selectlist}{itemize}{2}
\setlist[selectlist]{label=$\square$,leftmargin=*,noitemsep,topsep=0pt}

\usepackage{lmodern}

\usepackage[resetlabels,labeled]{multibib}
\newcites{URL}{Resources - URLs}

\PassOptionsToPackage{hyphens}{url}\usepackage{hyperref}
\hypersetup{
    colorlinks=true,
    linkcolor=blue,
    filecolor=magenta,      
    urlcolor=blue,
}

\makeatletter
\newcommand\footnoteref[1]{\protected@xdef\@thefnmark{\ref{#1}}\@footnotemark}
\makeatother

\titleformat{\section}[block]{\hspace{1em}\bfseries}{\thesection.}{0.5em}{} 
\titleformat{\subsection}[block]{\hspace{1em}}{\thesubsection}{0.5em}{}

\begin{document}
\begin{flushleft}

\setlength{\parindent}{0pt}
\setlength{\parskip}{10pt}

\textbf{LAMSkyCam: A Low-cost and Miniature Ground-based Sky Camera}

\textbf{Authors}\\
Mayank Jain$^{a,b}$~\footnote{\label{eqContrib}Authors contributed equally.}, Vishal Singh Sengar$^{c}$~\footnoteref{eqContrib}, Isabella Gollini$^{d}$, Michela Bertolotto$^{b}$, Gavin McArdle$^{b}$, and Soumyabrata Dev$^{a,b}$

\textbf{Affiliations}\\
\textit{$^a$~The ADAPT SFI Research Centre, Ireland\\
$^b$~School of Computer Science, University College Dublin, Ireland\\
$^c$~Indraprastha Institute of Information Technology, Delhi, India\\
$^d$~School of Mathematics and Statistics, University College Dublin, Ireland
}

\textbf{Corresponding author’s email address}\\
\textit{soumyabrata.dev@ucd.ie}

\textbf{Abstract}\\
Ground-based sky imagers (GSIs) are increasingly becoming popular amongst the remote sensing analysts. This is because such imagers offer fantastic alternatives to satellite measurements for the purpose of earth observations. In this paper, we propose an extremely low-cost and miniature ground-based sky camera for atmospheric study. Built using $3$D printed components and off-the-shelf components, our sky camera is lightweight and robust for use in diverse climatic conditions. With a $63^{\circ}$ field of view angle, the camera captures high resolution sky/cloud images for both day and night times at $5$ minute intervals. The camera is designed to be be mounted on a pole-like architecture and with its compact form, it can be installed at any location without requiring any change in the existing infrastructure. For remote areas, the camera has also a local backup facility from which data can be easily accessed manually. We have open-sourced the hardware design of our sky camera, and therefore researchers can easily manufacture and deploy these cameras for their respective use cases. 

\textbf{Keywords}\\ \textit{Remote Sensing, Raspberry Pi, $3$D-Printing, Cloud Images}

\newpage
\textbf{Specifications table}\\
\vskip 0.2cm
\tabulinesep=1ex
\begin{tabu} to \linewidth {|X|X[3,l]|}
\hline  \textbf{Hardware name} & LAMSkyCam: A Low-cost And Miniature Ground-based Sky Camera
  \\
  \hline \textbf{Subject area} & %
  \textit{Environmental, planetary and agricultural sciences}
  \\
  \hline \textbf{Hardware type} &
  \textit{Imaging tools}
  \\
\hline \textbf{Closest commercial analog} &
  TSI$-880$ Total Sky Imager~\cite{long2001total} by Yankee Environmental Systems, Massachusetts, USA.
  The proposed hardware reduces the cost drastically while providing images in much higher-resolution ($4056\times3040$) of both day and night sky as compared to $352\times288$ resolution daytime only images from the commercial counterpart.
  \\
\hline \textbf{Open source license} &
  \href{https://cern-ohl.web.cern.ch/}{CERN Open Hardware Licence (OHL)}
  \\
\hline \textbf{Cost of hardware} &
  USD $285$ (approximately)
  \\
\hline \textbf{Source file repository} & 
  \url{https://doi.org/10.17632/5fth8h9sdh.1}
  \\
\hline
\end{tabu}
\end{flushleft}
\newpage
\section{Hardware in context}
With the recent advancements in hardware advancements and higher computing power, there has been a massive interests amongst the researchers in using diverse sensor equipment. These sensor equipment are extensively used by earth observation researchers in understanding the various events in the atmosphere. Traditionally, events in the atmosphere are performed via satellite and air-borne images. However, these images suffer from the inherent drawback of low temporal resolution and low spatial resolution. Therefore, images obtained from ground-based sky imagers (GSIs) provide us a good alternative. In this paper, we propose an extremely low-cost, ground-based, miniature ground-based camera. We acronym our proposed GSI hardware as LAMSkyCam: Low-cost And Miniature ground-based Sky Camera. Our sky camera is extremely low-cost and is sourced from 3D-printed materials. This renders our hardware light-weight and low-cost, and enables researchers to deploy multiple sky cameras across an area. Our proposed sky camera can be used in a variety of applications, including atmospheric study~\cite{dev2019cloudsegnet,dev2015multi}, satellite communication~\cite{dev2017nighttime,dev2016color,manandhar2017correlating}, aviation~\cite{mandel1975early}, weather research~\cite{manandhar2018systematic,dev2017cloud}, solar and renewable energy~\cite{dev2019estimating,dev2016estimation}, cloud type recognition~\cite{dev2015categorization}, detecting rainfall onset~\cite{dev2016detecting}, and atmospheric pollutant analysis~\cite{skidmore2011all}. 

In the literature, several models of ground-based sky cameras are introduced for industrial and research uses. However, these imagers are expensive, used proprietary hardware, and were inflexible in its operation. TSI-880 manufactured by manufactured by Yankee Environmental Systems (YES) Incorporated, based in Turner Falls, Massachusetts is one of the pioneering models that was instituted~\cite{long2001total}. The TSI-880 was also extensively used by several research groups for the purpose of weather monitoring and atmospheric analysis. However, one of the main drawbacks of TSI-880 was its low image capturing resolution and high-cost at US\$30,000. Reuniwatt is another such enterprise which is engaged in manufacturing ground-based sky imagers~\cite{bergler_2019}. Their highly advanced systems can provide continuous streams of cloud images for both day and night times along with solar irradiance readings using a pyranometer. The only downside is their high costs. ASI-16 All Sky Imager from EKO Instruments Europe B.V.~\cite{po2018advanced} is another commercially available unit with similar characteristics. Additionally, other sky cameras were custom-built for diverse applications. Richardson~\textit{et~al.}~\cite{richardson2017low} designed an all-sky imager for tracking clouds and solar irradiance forecasting. Another high-dynamic-range ground-based camera was developed by Urquhart~\textit{et~al.}~\cite{urquhart2015development} for solar-power forecasting applications. Kazantzidis~\textit{et~al.} used ground-based whole sky imagers for cloud detection and cloud type recognition~\cite{kazantzidis2012cloud}. Long~\textit{et~al.}~\cite{long2006retrieving} used ground-based sky cameras for retrieving cloud characteristics. Recently, Dev~\textit{et~al.}~\cite{dev2014wahrsis} designed the first low-cost camera at US\$2,500 using off-the-shelf components~\cite{godrich2017students}, that proved the flexibility of using off-the-shelf components for designing sky cameras. It captured images at high image resolutions and used mechanical sun-blocker for reducing the glare of circumsolar region. The mechanical sun-blocker was removed in a later version of the camera~\cite{dev2015design}, and it used high-dynamic-range imaging~\cite{dev2018high} to reduce the sun glare in the captured sky/cloud image. More recent studies have reinforced that the requirement of sun tracking system can be nullified with wide field of view lens, better camera sensors and optimal post processing of captured images~\cite{fa2019development}.

Inspired from all the previous designs and with an aim to make an extremely low-cost sky imager, in the previous version of this paper~\cite{jain2021wsi}, we reduced the cost of our sky camera, paving its path for multiple camera installations in the chosen geographical study area. Such multiple installations are necessary for estimating important parameters like cloud base height using stereo-vision~\cite{po2018advanced,savoy2015cloud}. The earlier prototype further evolved into the current LAMSkyCam model with its lower cost, miniature size and robust design to handle extreme weather conditions.

\section{Hardware description}

\begin{figure}[!ht]
    \centering
    \includegraphics[trim={0 0 0 0},clip,width=0.8\textwidth]{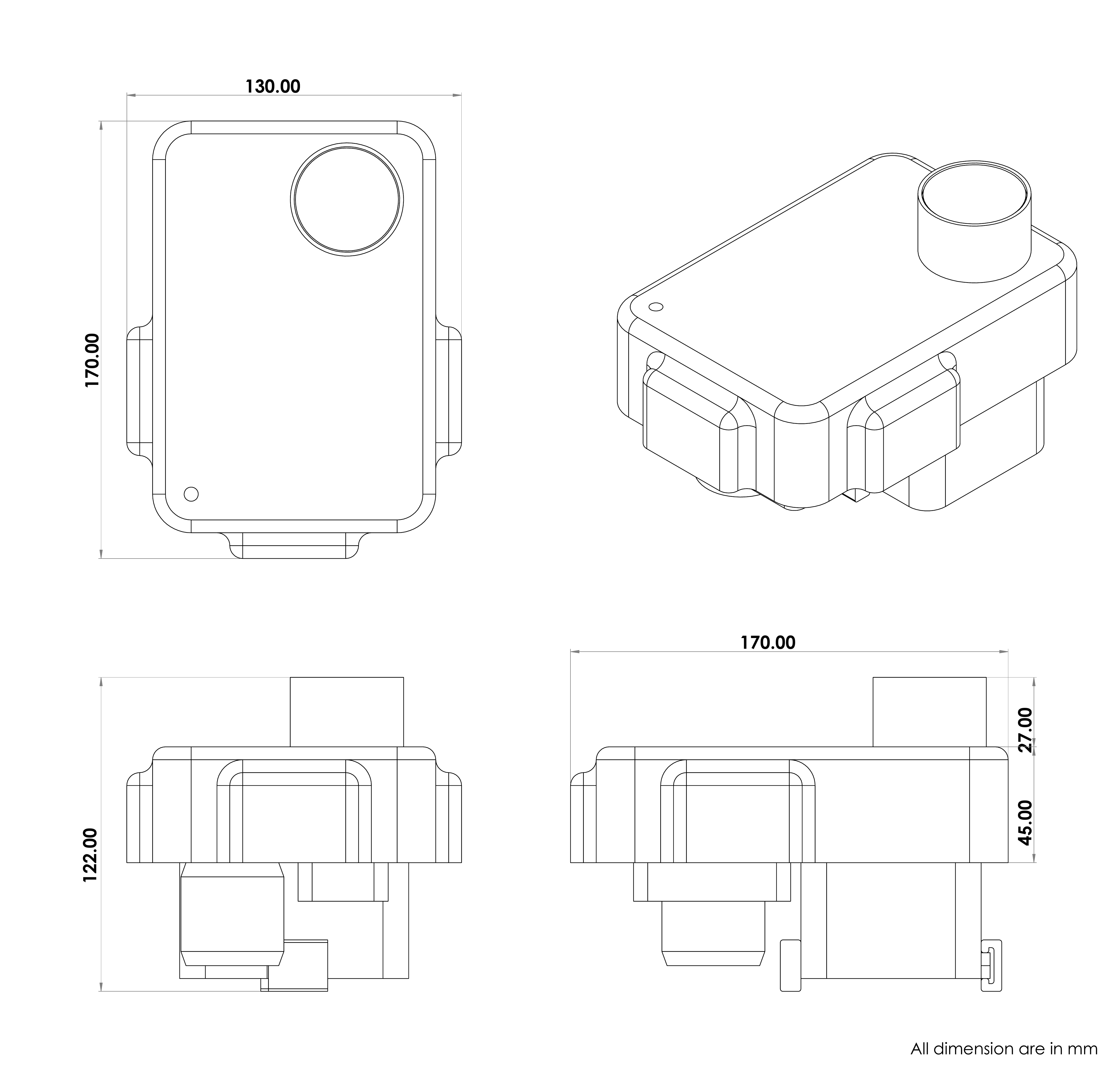}
    \caption{Dimensions of the designed GSI. All dimensions are in $mm$.}
    \label{fig:dimensionalView}
\end{figure}

\begin{figure}[!ht]
\centering
\begin{minipage}{.49\textwidth}
  \centering
  \includegraphics[trim={0 150 0 0},clip,width=0.95\linewidth,keepaspectratio]{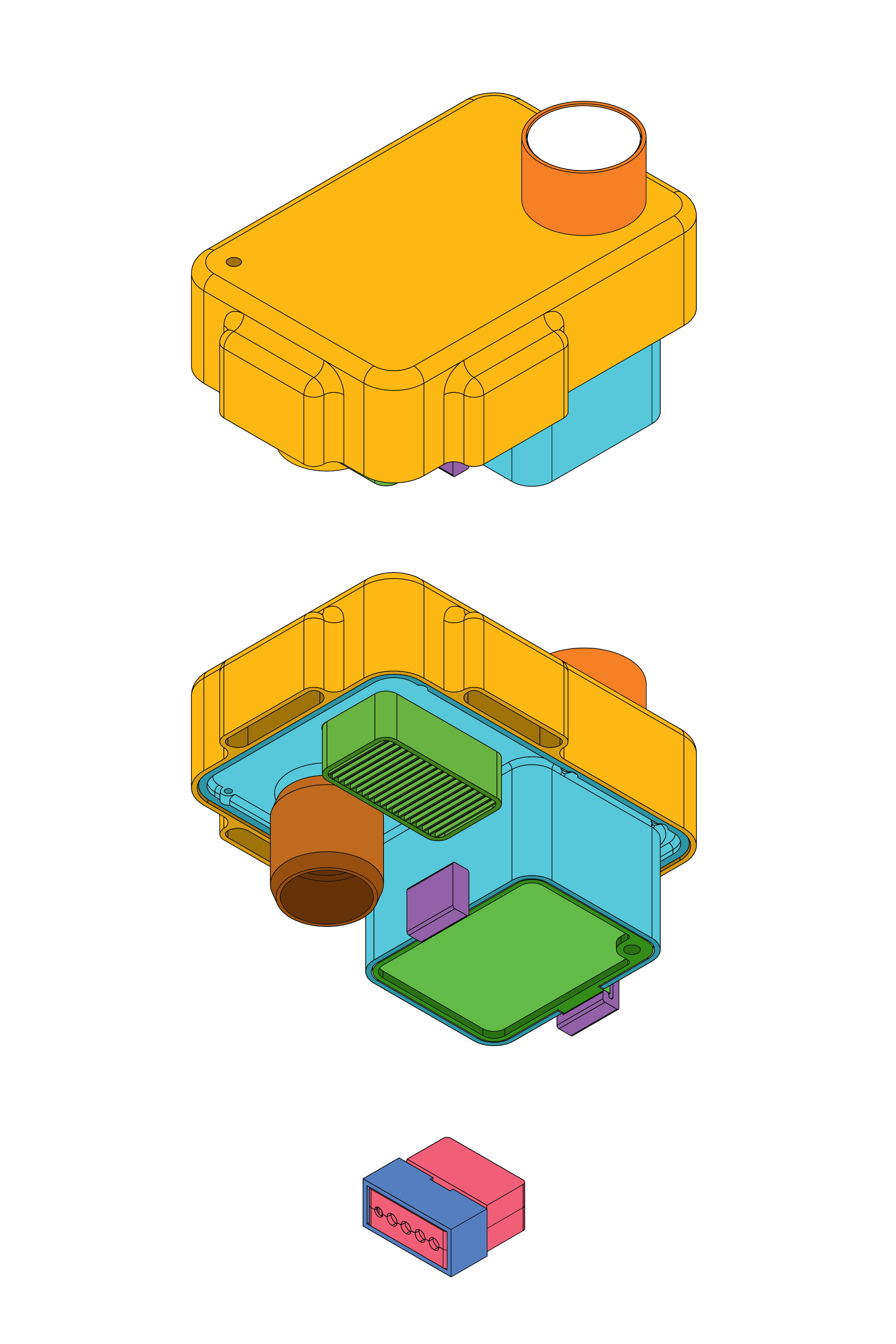}
  \caption{$3$D Perspective view of the designed GSI while showing the angled top and bottom views.}
  \label{fig:GSIcoloredDrawing}
\end{minipage}\hfill%
\begin{minipage}{.49\textwidth}
  \centering
  \includegraphics[trim={0 150 0 0},clip,width=0.95\linewidth,keepaspectratio]{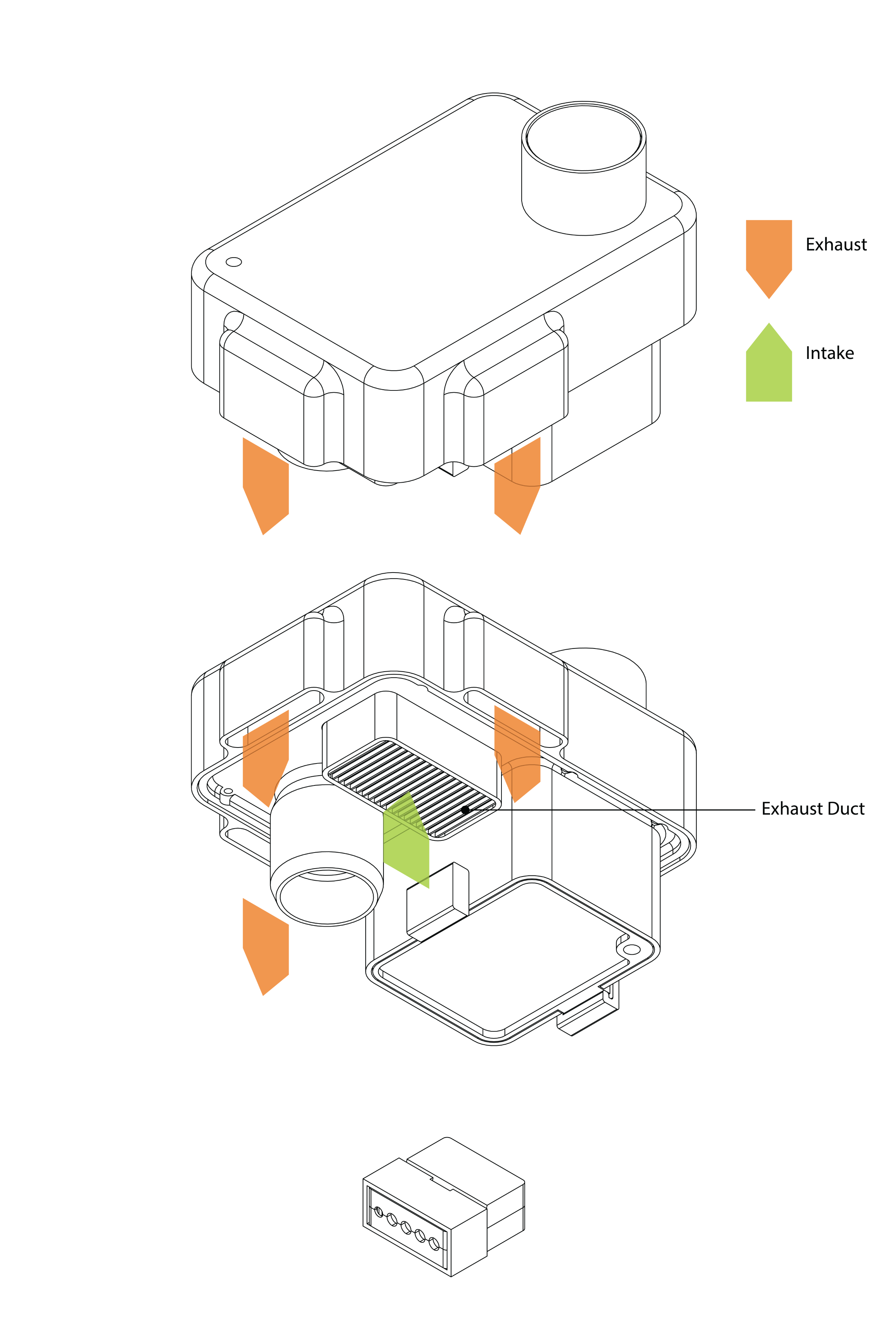}
  \caption{Exhaust and intake ducts of the GSI highlighted in the $3$D perspective view.}
  \label{fig:GSIexhaustANDinlet}
\end{minipage}
\end{figure}

The LAMSkyCam is ultra compact in size ($170\times130\times120mm$), extremely low-cost (USD $285$) and has the ability to capture images in both day and night at a high resolution ($4056\times3040$). Additionally, the designed GSI is capable of creating a local backup for installation in remote locations with no internet connection. This is possible due to the availability of low-cost and open-source electronics prototyping platform, i.e. Raspberry Pi~\cite{upton2012meet,pi2019raspberry}. Moreover, the platform comes with a vast amount of affordable electronic peripherals which come handy in many electronics prototyping applications. In this case, the Raspberry Pi HQ camera is used with a wide angle lens to construct the GSI. A dimensional view of the designed GSI is shown in the Figure~\ref{fig:dimensionalView}. Furthermore, the complete perspective view of the GSI is shown in Figure~\ref{fig:GSIcoloredDrawing}.

\begin{figure}[!htb]
    \centering
    \includegraphics[width=0.8\textwidth]{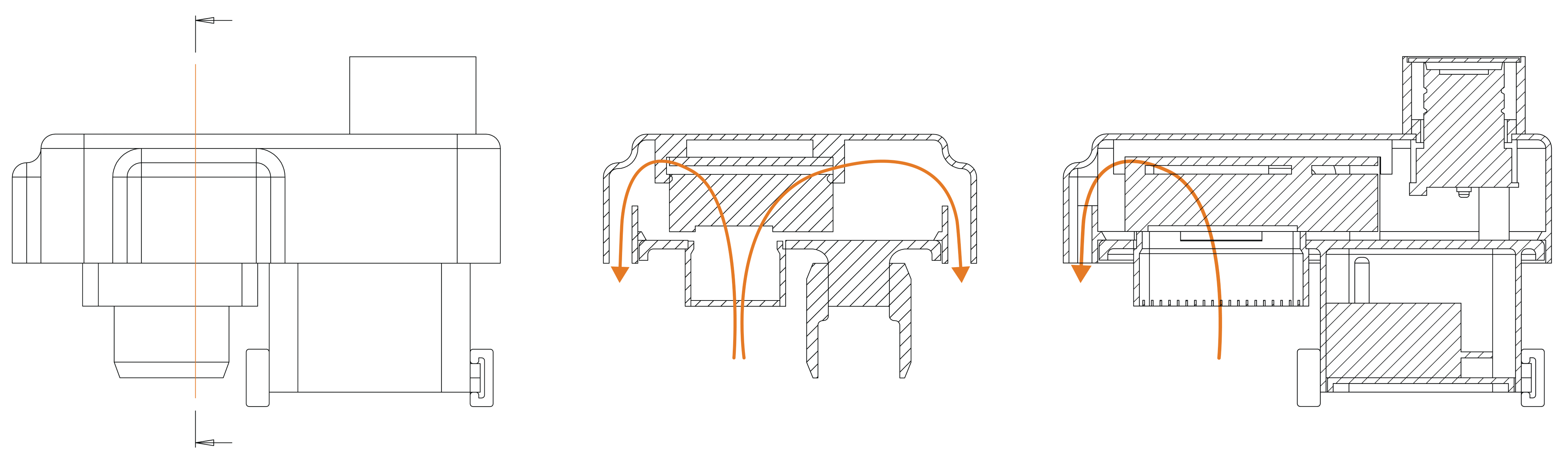}
    \caption{Air and heat circulation explained through the cross-section diagram of the designed GSI}
    \label{fig:airCirculation}
\end{figure}

With the help of the 3D-printing technology it is easier to make a design for the GSI which is water-proof. However, a bigger challenge is to get rid of the inside heat which is being generated by the electronic components (especially the Raspberry Pi) while keeping the overall GSI compact and still water-proof. One possibility is to provide a cavity near the base of the camera, letting the electronics completely exposed, which has been the case with some of the previous sky imagers~\cite{dev2014wahrsis,jain2021wsi}. Exposing the base leads to dust coming inside the camera, thereby opening up the possibility for corrosion and electronic malfunction. To resolve this issue, the Raspberry Pi is first covered with a metallic heat sink like armor from top and bottom. The heat sink has to small fans on one side which take in the air from the surroundings to cool the heat sink. This area of the armor with the fans is then connected to an exhaust duct which fits very closely to the armor itself making it very difficult for the external dust particles to enter the GSI's internal circuitry. While this takes care of the air inlet, heat outlets are provided by creating separate inverted and thin ducts on the periphery of the GSI (\textit{cf.} Figure~\ref{fig:GSIexhaustANDinlet}). While the inverted orientation doesn't allow rain to enter the GSI internals, the sleek design traps most of the dust particles by loosing their momentum before they reach the internal opening. Despite all the efforts, in case, if some dust particles manage to reach the inside of the GSI, the compact design ensures that there is no wind turbulence inside and the electronics is mounted upside down negating the presence of the dust particles that got trapped inside. Figure~\ref{fig:airCirculation} explains the air and heat circulation system of the GSI.

\begin{figure}[!htb]
    \centering
    \includegraphics[width=0.45\textwidth]{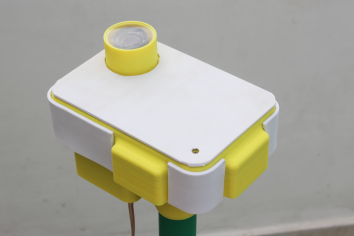}\hfill
    \includegraphics[width=0.45\textwidth]{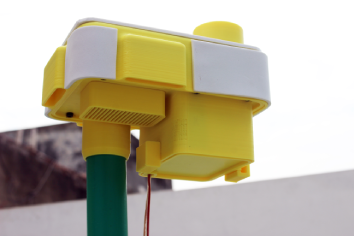}\\
    \makebox[0.45\textwidth][c]{\texttt{(a)}}\hfill
    \makebox[0.45\textwidth][c]{\texttt{(b)}}\\
    \caption{$3$D perspective views of the completed GSI in operation: \texttt{(a)} - angled top-left; \texttt{(b)} - angled bottom-right}
    \label{fig:closeUpGSI}
\end{figure}

With the internally generated heat taken care of, the GSI will work under the normal operating temperature range of the Raspberry Pi and the Raspberry Pi HQ camera, i.e. $0-50^{\circ}C$. To make the camera robust towards lower and higher temperatures than the aforementioned range, additional heating and cooling systems will be required to be installed respectively, along with a temperature and humidity monitoring sensor. Without including extra hardware, heat insulating foam sheets were attached to the camera in order to slightly increase the temperature tolerance, especially in higher temperatures. The addition of this insulating sheet has been noted to reduce the temperature inside the GSI by up to $2^{\circ}C$. Figure~\ref{fig:closeUpGSI} shows the finally constructed GSI when viewed up close from the angled top-left and angled bottom-right perspectives.

The camera features a $63^{\circ}$ field of view wide angle camera lens with $6mm$ focal length. Given the minimum height of clouds from ground is around $2km$~\cite{zhang2016observation}, this means that the lens can cover a total distance of $tan\left(\frac{63}{2}\right)\times2 = 1.22km$ in all $4$ directions. However, the actual camera sensor captures images in a rectangular shape (i.e. $4056\times3040$) instead of a square shape. Assuming that the sensor crops (or looses) the information at the shorter side, the distance covered by the complete GSI in the shorter capture side will be $=2\times\frac{3040}{4056}\times1.22 = 1.84km$. Since the cost of the GSI is very low, one can install multiple GSIs at around $1.84-2.44km$ distance from each other to cover more area.

Additionally, the LAMSkyCam is equipped with an anti-reflective and UR filter lens cover which helps to get sharper images and reduces the effect of sun glare in critical photographs. Moreover, since such cameras are generally placed in remote locations with no internet connectivity, the GSI features an external storage device which can contain data for more than $3$ months and the backup can then be manually obtained. Furthermore, the miniature size of the camera and the provision of separate pole-like mounting, the camera can be placed at any location without disturbing the existing infrastructure.

The salient features and use-cases of the LAMSkyCam can now be summarised as follows:
\begin{itemize}
    \item The designed GSI can capture sky/cloud images during both day and night.
    \item These images are of a high resolution ($4056\times3040$).
    \item The compact size of the GSI and its mountability on any pole makes it easier to install.
    \item The camera can operate in the temperature range of $0-50^{\circ}C$.
    \item The camera hosts a local backup facility with capacity of data from more than $3$ months.
    \item Images from the GSI can be used for various meteorological studies, specifically cloud movements at high temporal and spatial resolution.
\end{itemize}

\section{Design files summary}\label{sec:DesignFiles}
\vskip 0.1cm
\tabulinesep=1ex
\begin{longtabu} to \linewidth {|X|X|X[1.1,1]|X[1.1,1]|} 
\hline
\textbf{Design filename} & \textbf{File type} & \textbf{Open source license} & \textbf{Location of the file} \\\hline
Part $\# 01$ & STEP (CAD), STL & \href{https://cern-ohl.web.cern.ch/}{CERN OHL} & \citeURL{MendeleyDOI, MainBody, MainBodySTL}  \\\hline
Part $\# 02$ & STEP (CAD), STL & \href{https://cern-ohl.web.cern.ch/}{CERN OHL} & \citeURL{MendeleyDOI, MainBodyCover, MainBodyCoverSTL} \\\hline
Part $\# 03$ & STEP (CAD), STL & \href{https://cern-ohl.web.cern.ch/}{CERN OHL} & \citeURL{MendeleyDOI, Duct, DuctSTL} \\\hline
Part $\# 04$ & STEP (CAD), STL & \href{https://cern-ohl.web.cern.ch/}{CERN OHL} & \citeURL{MendeleyDOI, Lid, LidSTL} \\\hline
Part $\# 05$ & STEP (CAD), STL & \href{https://cern-ohl.web.cern.ch/}{CERN OHL} & \citeURL{MendeleyDOI, LidRetainer, LidRetainerSTL} \\\hline
Part $\# 06$ & STEP (CAD), STL & \href{https://cern-ohl.web.cern.ch/}{CERN OHL} & \citeURL{MendeleyDOI, LensHoodMount, LensHoodMountSTL} \\\hline
Part $\# 07$ & STEP (CAD), STL & \href{https://cern-ohl.web.cern.ch/}{CERN OHL} & \citeURL{MendeleyDOI, LensHood, LensHoodSTL} \\\hline
Part $\# 08$ & STEP (CAD), STL & \href{https://cern-ohl.web.cern.ch/}{CERN OHL} & \citeURL{MendeleyDOI, Remote, RemoteSTL} \\\hline
Part $\# 09$ & STEP (CAD), STL & \href{https://cern-ohl.web.cern.ch/}{CERN OHL} & \citeURL{MendeleyDOI, RemoteMirror, RemoteMirrorSTL} \\\hline
Part $\# 10$ & STEP (CAD), STL & \href{https://cern-ohl.web.cern.ch/}{CERN OHL} & \citeURL{MendeleyDOI, RemoteLock, RemoteLockSTL} \\\hline
Part $\# 11$ & STEP (CAD), STL & \href{https://cern-ohl.web.cern.ch/}{CERN OHL} & \citeURL{MendeleyDOI, MountSocket, MountSocketSTL} \\\hline
Part $\# 12$ & STEP (CAD), STL & \href{https://cern-ohl.web.cern.ch/}{CERN OHL} & \citeURL{MendeleyDOI, Tripod, TripodSTL} \\\hline
Part $\# 13$ & PDF & \href{https://cern-ohl.web.cern.ch/}{CERN OHL} & \citeURL{MendeleyDOI, Insulation} \\\hline
\end{longtabu}

\vskip 0.3cm
\noindent
\begin{itemize}
\item[$\bullet$]\textbf{Part $\# 01$} is main body which is designed to be printed as a single part that can be printed without any support. The main body contains an air duct that helps the exhaust air to escape.
\item[$\bullet$]\textbf{Part $\# 02$} is main body cover which is attached to the main body by $6$ $M3$ $10mm$ bolts. The cover contains storage for storage device, remote, and power adapter. The cover also contains mounting points for duct vent.
\item[$\bullet$]\textbf{Part $\# 03$} is duct which is designed to be printed separately, even though its a part of the main body cover. This ensures that the rest of the cover can be $3$D printed as a single part. The duct needs to be attached to the cover using a super glue cover after securing the cover using M3 bolts. It provides a channel for the exhaust fans on Raspberry Pi to intake air from outside.
\item[$\bullet$]\textbf{Part $\# 04$} is lid which is designed for ease of access to the storage device, adapter, and the remote. The lid has a port for power cable, and it helps to protect the peripherals from outside weather conditions.
\item[$\bullet$]\textbf{Part $\# 05$} is the lid retainer which is designed to hold the lid steady with the Main Body Cover. $2$ such retainers hold the lid from either side. Lid can then be detached by sliding the retainer.
\item[$\bullet$]\textbf{Part $\# 06$} is the lens hood mount which is also designed to be printed separately so that the main body can be $3$D printed without support. This needs to be fixed with super glue to the main body.
\item[$\bullet$]\textbf{Part $\# 07$} is lens hood which is designed so that it can be detached from the main body in case the lens is required to be cleaned or serviced.
\item[$\bullet$]\textbf{Part $\# 08$} is first half of remote. Remote can be used to safely mount and eject the storage device and also show the status of the operation.
\item[$\bullet$]\textbf{Part $\# 09$} is the second half of the Remote and the mirror of the first half. Separate halves make it easy to assemble $4$ LEDs and a push button inside the remote in minimal space.
\item[$\bullet$]\textbf{Part $\# 10$} is the remote lock which keeps the both halves of the remote joined together and give the remote a complete look.
\item[$\bullet$]\textbf{Part $\# 11$} is mount socket which is designed to be printed separately from the cover so that the diameter can be changed according to the mounting requirements.
\item[$\bullet$]\textbf{Part $\# 12$} is tripod which is currently designed as a joint for $4$ plastic pipes. The tripod can be edited for different diameters of pipes.
\item[$\bullet$]\textbf{Part $\# 13$} is the design file contains the measurements and design specifications which are required to cut the insulating sheet. These insulating sheet cut-outs are required to be pasted on top of the final GSI to protect the same from outside heat.
\end{itemize}

\section{Bill of materials summary}\label{sec:BOM}
\vskip 0.2cm
\tabulinesep=1ex
\noindent
\begin{longtabu} to \linewidth {|X[0.6,1]|X[1.3,1]|X[0.5,1]|X[0.6,1]|X[0.65,1]|X[0.6,1]|X[0.6,1]|}
\hline
\textbf{Designator} & \textbf{Component} & \textbf{Number} & \textbf{Cost/Unit (USD)} & \textbf{Total Cost (USD)} & \textbf{Source of materials} & \textbf{Material type} \\\hline

Part $\# 01$ & Main Body
& $1$ ($\times165g$) & $27.2$ ($/kg$) & $4.49$ & \citeURL{PLA} & Polymer \\\hline

Part $\# 02$ & Main Body Cover & $1$ ($\times76g$) & $27.2$ ($/kg$) & $2.07$ & \citeURL{PLA} & Polymer \\\hline

Part $\# 03$ & Duct & $1$ ($\times13g$) & $27.2$ ($/kg$) & $0.35$ & \citeURL{PLA} & Polymer \\\hline

Part $\# 04$ & Lid & $1$ ($\times15g$) & $27.2$ ($/kg$) & $0.41$ & \citeURL{PLA} & Polymer \\\hline

Part $\# 05$ & Lid Retainer & $2$ ($\times6g$) & $27.2$ ($/kg$) & $0.33$ & \citeURL{PLA} & Polymer \\\hline

Part $\# 06$ & Lens Hood Mount & $1$ ($\times3g$) & $27.2$ ($/kg$) & $0.08$ & \citeURL{PLA} & Polymer \\\hline

Part $\# 07$ & Lens Hood & $1$ ($\times9g$) & $27.2$ ($/kg$) & $0.24$ & \citeURL{PLA} & Polymer \\\hline

Part $\# 08$ & Remote $1^{st}$ Half & $1$ ($\times8g$) & $27.2$ ($/kg$) & $0.22$ & \citeURL{PLA} & Polymer \\\hline

Part $\# 09$ & Remote $2^{nd}$ Half & $1$ ($\times8g$) & $27.2$ ($/kg$) & $0.22$ & \citeURL{PLA} & Polymer \\\hline

Part $\# 10$ & Remote Lock & $1$ ($\times6g$) & $27.2$ ($/kg$) & $0.16$ & \citeURL{PLA} & Polymer \\\hline

Part $\# 11$ & Mount Socket & $1$ ($\times26g$) & $27.2$ ($/kg$) & $0.71$ & \citeURL{PLA} & Polymer \\\hline

Part $\# 12$ & Tripod & $1$ ($\times50g$) & $27.2$ ($/kg$) & $1.36$ & \citeURL{PLA} & Polymer \\\hline

Part $\# 13$ & Insulation Sheet ($500mm\times200mm\times3mm$) & $1$ & $3.80$ ($/10$) & $0.38$ & \citeURL{InsulationSheet} & Polymer \\\hline

Part $\# 14$ & Raspberry Pi $4$ Model-B (w/$4GB$ RAM) & $1$ & $67.89$ & $67.89$ & \citeURL{RaspPI} & Other \\\hline

Part $\# 15$ & USB-C $15.3W$ Power Supply & $1$ & $10.19$ & $10.19$ & \citeURL{USBcharger} & Other \\\hline

Part $\# 16$ & Raspberry Pi HQ Camera with Interchangeable Lens Base & $1$ & $67.33$ & $67.33$ & \citeURL{HQcamera} & Other \\\hline

Part $\# 17$ & Raspberry Pi $6mm$ Wide Angle Lens & $1$ & $31.96$ & $31.96$ & \citeURL{WideLens} & Other \\\hline

Part $\# 18$ & DS$3231$ RTC Module & $1$ & $4.75$ & $4.75$ & \citeURL{RTC} & Other \\\hline

Part $\# 19$ & Micro ($\mu$SDXC) $64GB$ Class $10$ Memory Card & $1$ & $10.05$ & $10.05$ & \citeURL{uSDcard} & Other \\\hline

Part $\# 20$ & USB $3.0$ $256GB$ Solid State Flash Drive & $1$ & $35.03$ & $35.03$ & \citeURL{PenDrive} & Other \\\hline

Part $\# 21$ & Raspberry Pi $4$ Metal Aluminium Alloy-Based Cooling Heatsink Armor Case (w/Dual Fan) & $1$ & $19.03$ & $19.03$ & \citeURL{PiCase} & Other \\\hline

Part $\# 22$ & USB $3.0$ Male A to Female A Extension Cable ($30cm$) & $1$ & $4.07$ & $4.07$ & \citeURL{USBext} & Other \\\hline

Part $\# 23$ & CR$2032$ Coin Battery $3V$ & $1$ & $1.09$ & $1.09$ & \citeURL{CoinCell} & Other \\\hline

Part $\# 24$ & Light Dependent Resistance (LDR) & $1$ & $0.80$ ($/10$) & $0.08$ & \citeURL{LDR} & Other \\\hline

Part $\# 25$ & Light Emitting Diode (LED) & $4$ & $3.39$ ($/100$) & $0.14$ & \citeURL{LED} & Other \\\hline

Part $\# 26$ & $330\Omega$ Resistor & $4$ & $1.35$ ($/100$) & $0.05$ & \citeURL{Resistor} & Other \\\hline

Part $\# 27$ & $1M\Omega$ Resistor & $1$ & $1.35$ ($/100$) & $0.01$ & \citeURL{Resistor} & Other \\\hline

Part $\# 28$ & $1\mu F$ Electrolytic Capacitor & $1$ & $1.07$ ($/40$) & $0.03$ & \citeURL{Capacitor} & Other \\\hline

Part $\# 29$ & Tactile Push Button Switch & $1$ & $0.80$ ($/20$) & $0.04$ & \citeURL{PushButton} & Other \\\hline

Part $\# 30$ & Female - Female Jumper Cables for Electronic Prototyping & $40$ & $0.67$ ($/40$) & $0.67$ & \citeURL{FtoFjumper} & Other \\\hline

Part $\# 31$ & $40mm$ Anti-Reflective UV-Cut Curved Glass Disc (Lens Cover) & $1$ & $9.52$ & $9.52$ & Locally Sourced & Inorganic \\\hline

Part $\# 32$ & PVC Pipe for Water Fitting ($3m$) & $1$ & $5.44$ & $5.44$ & Locally Sourced & Polymer \\\hline

\end{longtabu}
\vskip 0.2cm
\noindent
Apart from the components described above, some generic workshop materials like, super glue~\citeURL{SuperGlue}, $M3$ $10mm$ screw bolts~\citeURL{Bolts}, and heat shrink tubes were used to complete the fabrication of the GSI. Estimated cost of such components is around $3$ USD. Part~$\# 31$ was sourced from a local optician as most have access to plain curved glass discs which have anti-reflective and UV-filter coating. Part~$\# 32$ is the polyvinyl chloride (PVC) pipes that are used for water supply fittings in households. Any other component can also be used for the tripod, however, in that case Part~$\# 11-12$ needs to be remodelled according to the tripod replacement one is using. Also, note that the components were originally procured in India (INR). Hence, the costs mentioned above were post conversion from INR to USD at the approximated currency exchange rate of $1$ USD $=$ $73.5$ INR.

\section{Build instructions}

Complete body of the GSI is $3$D printed using a Fused Deposition Modeling (FDM) technology based printer. While the component names and description has been mentioned in the sections~\ref{sec:DesignFiles} and~\ref{sec:BOM}, following is the checklist of common workshop components/tools which were required to completely build the proposed design of GSI:
\begin{selectlist}
    \item $3$D printer
    \item Soldering wire and soldering station
    \item Sandpaper
    \item Metric allen key set
    \item M$3$ $10mm$ nuts and bolts (allen type)
    \item Heat shrink tube ($3mm$, and $6mm$)
    \item Hot air gun
    \item Fast and strong adhesive (super glue)
    \item Epoxy glue
    \item Heat resistant glue
    \item Masking tape
    \item Electrical insulation tape
    \item Electrical AC power supply cable with AC plug attached to one end
\end{selectlist}
Post procuring the components and $3$D printing the design files (as mentioned in section~\ref{sec:DesignFiles}), the first step is to assemble and prepare the electronic components. To setup the Raspberry Pi $4$ Model-B (Part~$\# 14$), the $64GB$ memory card (Part~$\# 19$) needs to be flashed for Raspbian OS. A complete tutorial for the same can be referenced from the Raspberry Pi official website~\citeURL{RPiGettingStarted}. To save additional trouble of installing softwares at a later stage, we can simply install the `Raspberry Pi OS with desktop and recommended software' from the list of operating system (OS) images that have been provided on the official website of the Raspberry Pi~\citeURL{RPiOS}.

\begin{figure}[!ht]
    \centering
    \includegraphics[width=0.8\textwidth]{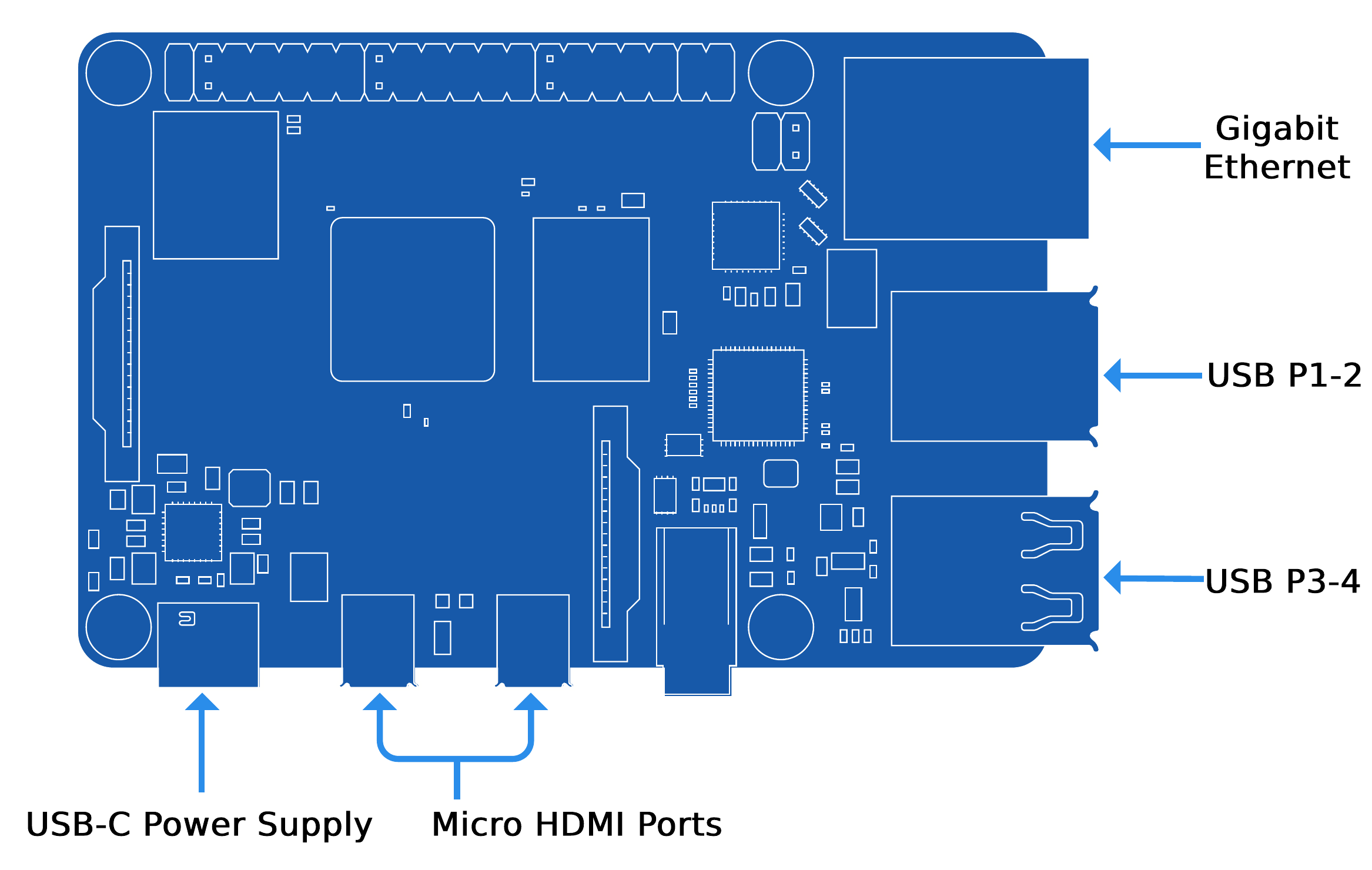}
    \caption{Blueprint of the Raspberry Pi $4$ Model-B which is being used in this work~\protect\citeURL{RPiBlueprint}.}
    \label{fig:RPiBlueprint}
\end{figure}

\begin{figure}[!ht]
    \centering
    \includegraphics[width=0.8\textwidth]{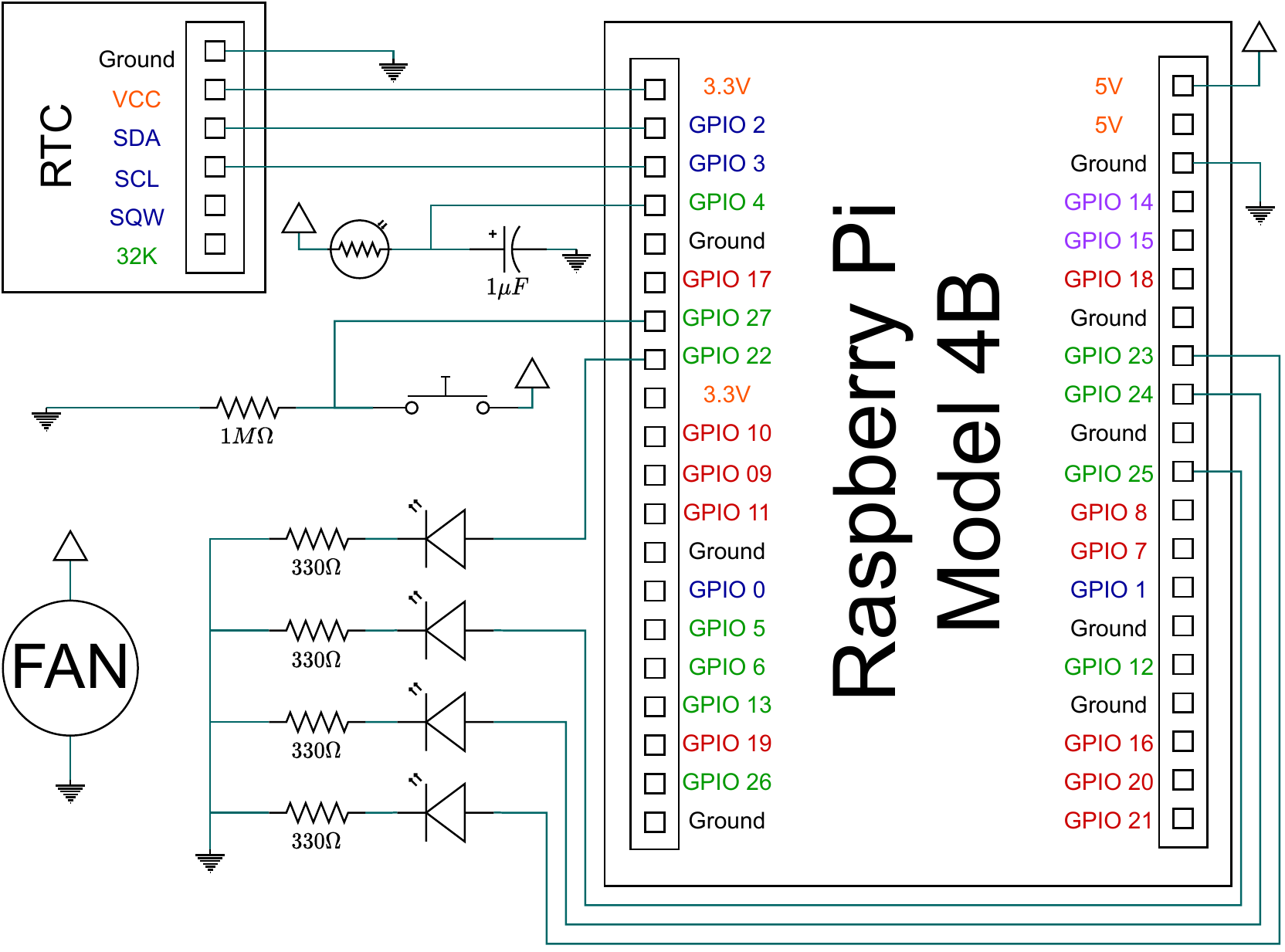}
    \caption{Electronic schematic of the GSI}
    \label{fig:circuitDiagram}
\end{figure}

Once the Raspbian OS is setup, the electronic components (Part~$\# 16$, Part~$\# 18$, Part~$\# 20-30$)  are connected to the Raspberry Pi. The heat sink armor casing for Raspberry Pi (Part~$\# 21$) is attached to the Raspberry Pi in the first step. Note that we must connect the ribbon connector cable of Raspberry Pi HQ camera (Part~$\# 16$) to the Raspberry Pi board before installing the armor casing as it cannot be done afterwards. Once done, the camera can then be attached to the other end of the ribbon connector cable. The USB $256GB$ flash drive (Part $\# 20$) is then connected to the upper USB port on the `USB P$3-4$' terminals depicted on the Raspberry Pi blueprint in the Figure~\ref{fig:RPiBlueprint} via the USB extension cable (Part~$\# 22$). The CR$2032$ coin battery (Part~$\# 23$) is then inserted into the DS$3231$ RTC module (Part~$\# 18$) for its proper functioning. Finally, the rest of the electronics is connected as per the schematic shown in the Figure~\ref{fig:circuitDiagram}. It is to be noted here that these connections are to be made in air without any breadboard or a printed circuit board (PCB) plate. This is because these components will later be directly attached to the different $3$D printed components to give things a finished and compact look. Here, we will isolate the wires going to the $2$ legs of the LDR ((Part~$\# 24$) as it will be attached to a separate part of the GSI chassis. The wires are doubled in length with the help of soldering such that the component can be routed properly. Similarly, the DS$3231$ RTC module connections are kept separately as it will fall in at another place in the GSI chassis. However, in this case, the wire lengths need not to be doubled. Finally, the rest of the electronic circuit components are soldered nearby to each other with jumper cables tripled in lengths as they will all go to the remote component of the GSI which needs to be extracted out of the camera body during operation. Heat shrink tubes and hot air gun is used as per the requirement for in air circuit stability. The hardware assembly steps that are discussed later in this section will provide a better picture of the setup, \textit{Note:} Please ensure safety while soldering and/or adding heat shrink tubes on the circuit wires since it might get tricky sometimes to solder components in air without any support of the PCB plate. Also, make sure that there are no cross-connections taking place once the setup is completed.

\begin{figure}[!ht]
    \centering
    \includegraphics[width=0.55\textwidth]{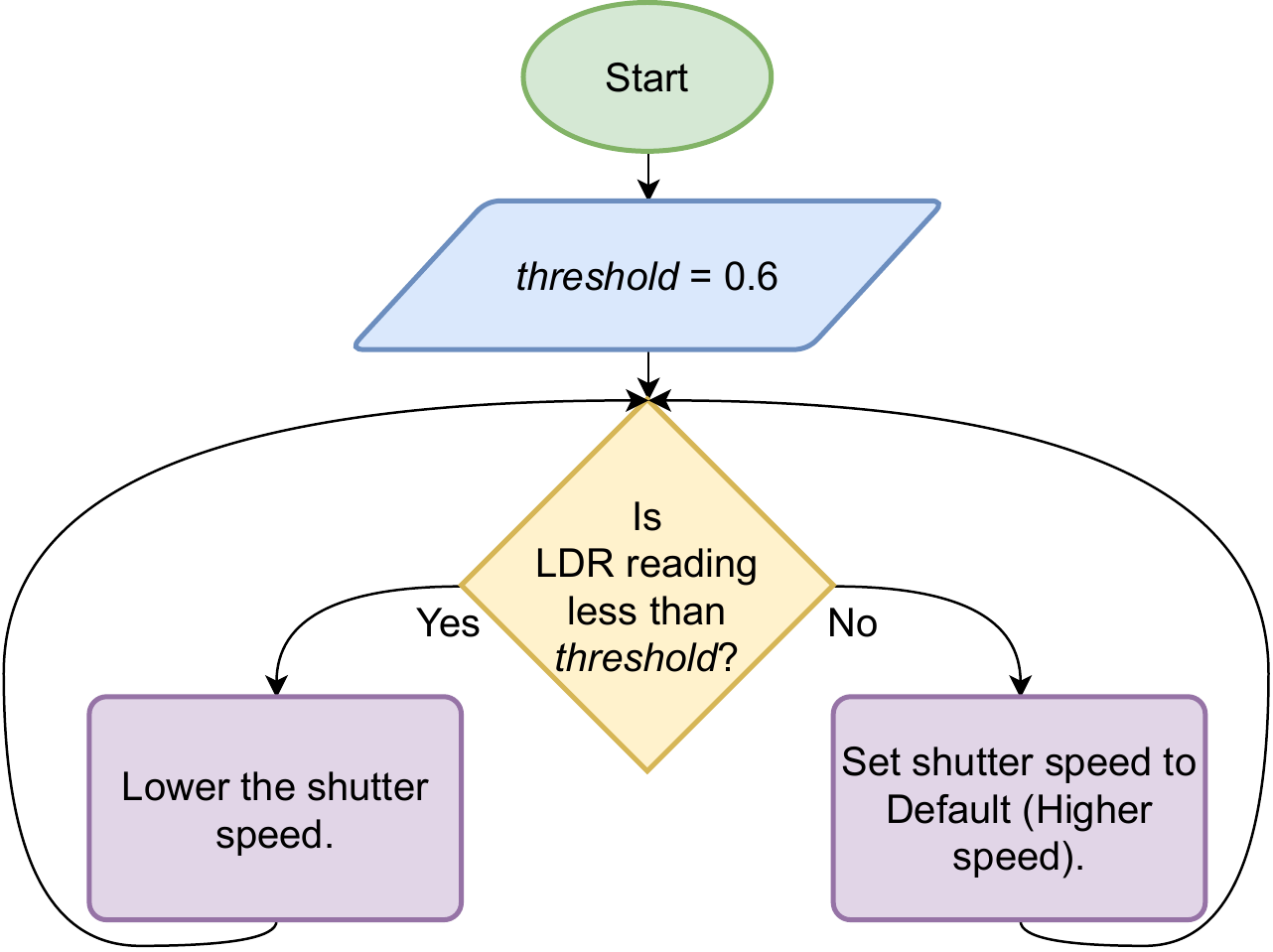}
    \caption{Flowchart depicting the camera shutter speed adjustment routine.}
    \label{fig:cameraCaptureFC}
\end{figure}

Next phase is to manage the software inside the Raspberry Pi\footnote{In the spirit of reproducible research, the code for the software described in this paper is shared at \url{https://github.com/jain15mayank/LAMSkyCam/tree/main/src}}. In this phase, the first task is to configure the DS$3231$ RTC module with the Raspberry Pi. Being one of the most popular RTC modules, one can easily find the installation instructions online~\citeURL{RTCconfig}. Next task is to automatically capture the images from the camera, A cron job~\cite{keller1999take,crontab} is written to capture the images from the camera at an interval of $5$ minutes. This makes the procedure fail-safe as the cron jobs are inherently managed by the operating system itself and hence can only fail in the case of complete system failure. On the other hand, a separate script adjusts the shutter speeds of the camera based on the LDR (Part $\# 24$) readings. The algorithm for this script is depicted in the form of a flowchart in Figure~\ref{fig:cameraCaptureFC}. Note that the value of \textit{threshold} variable in Figure~\ref{fig:cameraCaptureFC} must be adjusted depending on the procedure used to fetch and scale LDR readings (in this case, it was $0-1$) and the place of installation.

\begin{figure}[!htb]
    \centering
    \includegraphics[width=0.95\textwidth]{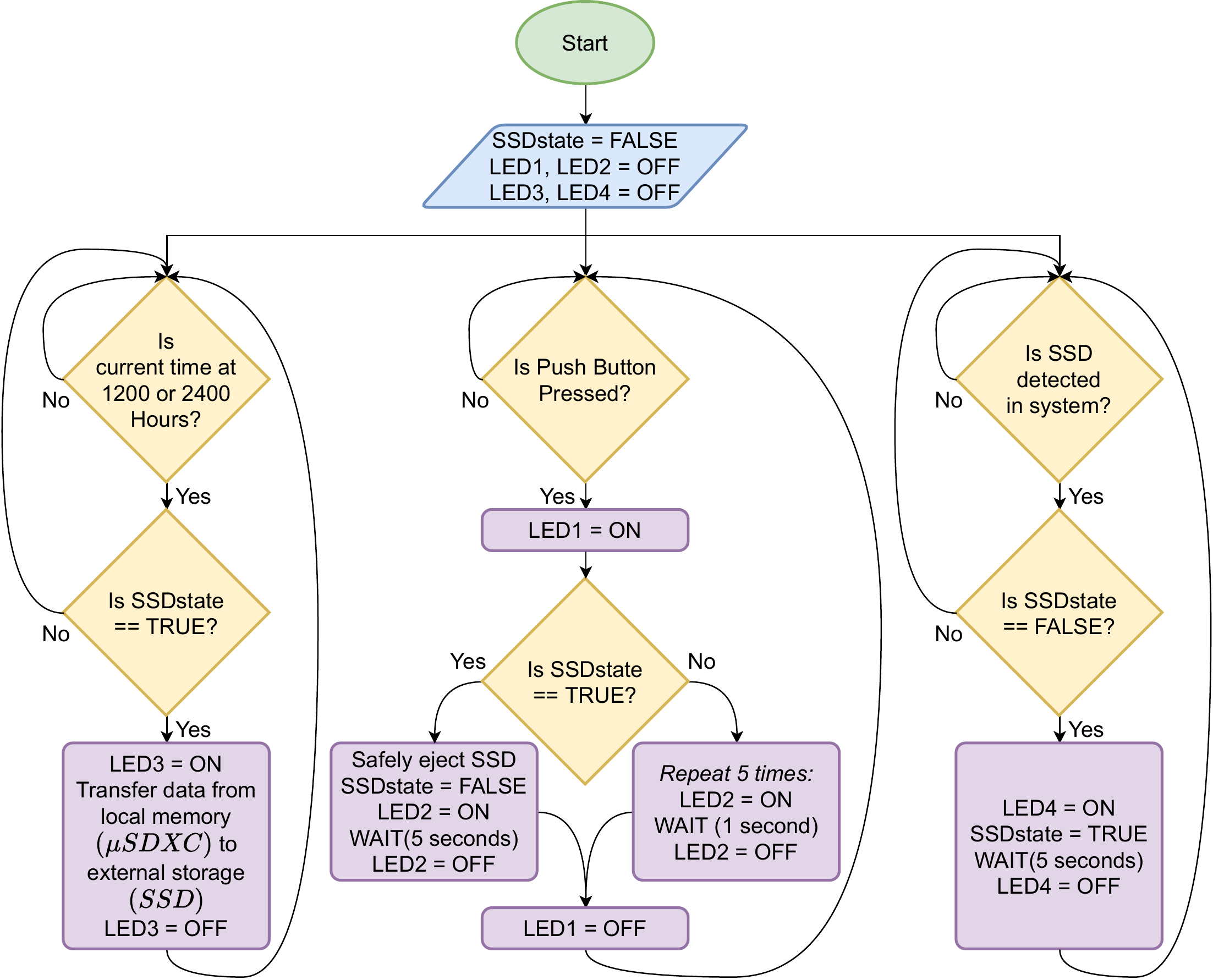}
    \caption{Flowchart depicting the SSD management routine.}
    \label{fig:SSDmanagementFC}
\end{figure}

\begin{figure}[!htb]
    \centering
    \includegraphics[width=0.3\textwidth]{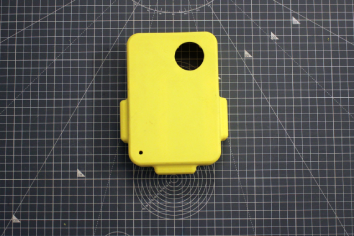}\hfill
    \includegraphics[width=0.3\textwidth]{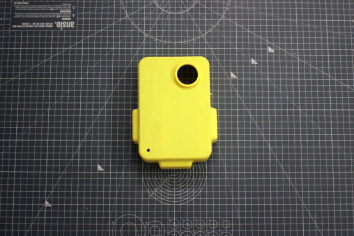}\hfill
    \includegraphics[width=0.3\textwidth]{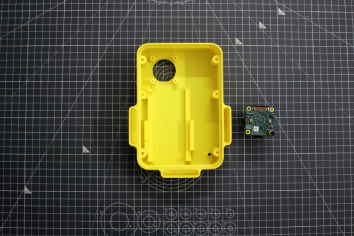}\\
    \makebox[0.3\textwidth][c]{\texttt{(a)}}\hfill
    \makebox[0.3\textwidth][c]{\texttt{(b)}}\hfill
    \makebox[0.3\textwidth][c]{\texttt{(c)}}\\\vskip 0.2cm
    \includegraphics[width=0.3\textwidth]{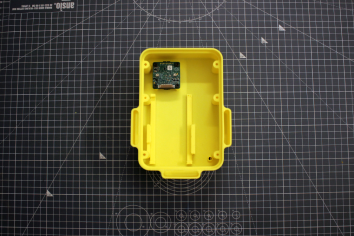}\hfill
    \includegraphics[width=0.3\textwidth]{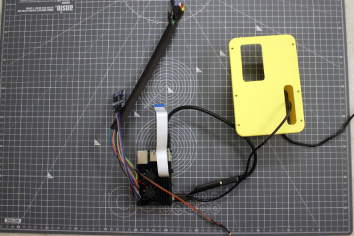}\hfill
    \includegraphics[width=0.3\textwidth]{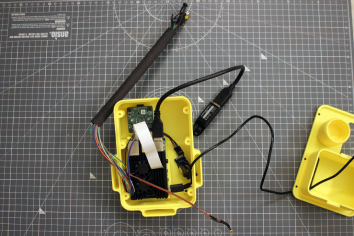}\\
    \makebox[0.3\textwidth][c]{\texttt{(d)}}\hfill
    \makebox[0.3\textwidth][c]{\texttt{(e)}}\hfill
    \makebox[0.3\textwidth][c]{\texttt{(f)}}\\\vskip 0.2cm
    \includegraphics[width=0.3\textwidth]{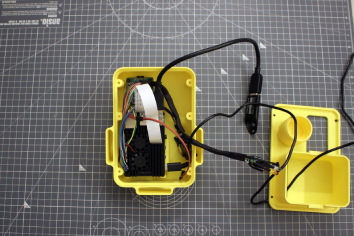}\hfill
    \includegraphics[width=0.3\textwidth]{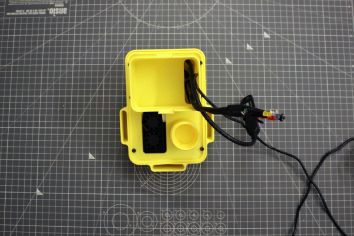}\hfill
    \includegraphics[width=0.3\textwidth]{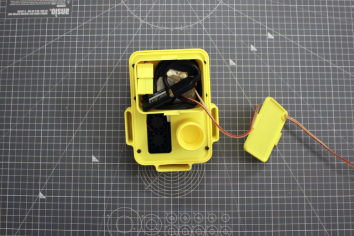}\\
    \makebox[0.3\textwidth][c]{\texttt{(g)}}\hfill
    \makebox[0.3\textwidth][c]{\texttt{(h)}}\hfill
    \makebox[0.3\textwidth][c]{\texttt{(i)}}\\\vskip 0.2cm
    \includegraphics[width=0.3\textwidth]{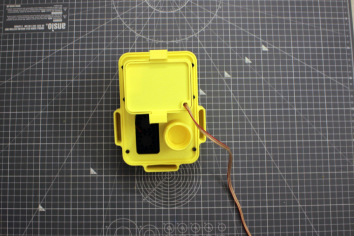}\hfill
    \includegraphics[width=0.3\textwidth]{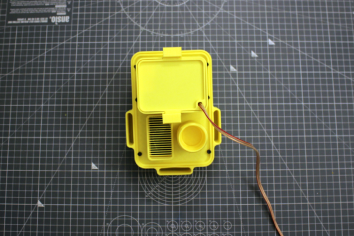}\hfill
    \includegraphics[width=0.3\textwidth]{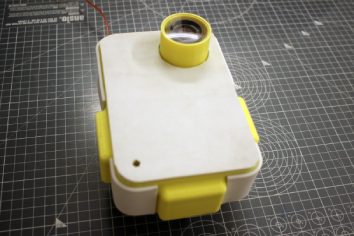}\\
    \makebox[0.3\textwidth][c]{\texttt{(j)}}\hfill
    \makebox[0.3\textwidth][c]{\texttt{(k)}}\hfill
    \makebox[0.3\textwidth][c]{\texttt{(l)}}\\\vskip 0.2cm
    \includegraphics[width=0.3\textwidth]{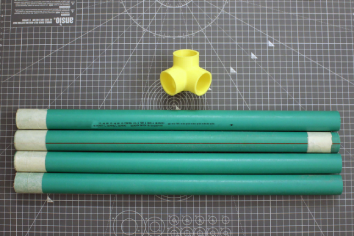}\hfill
    \includegraphics[width=0.3\textwidth]{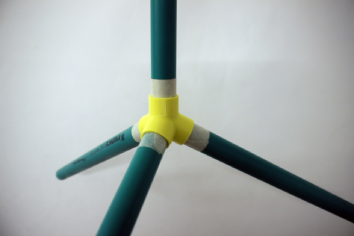}\hfill
    \includegraphics[width=0.3\textwidth]{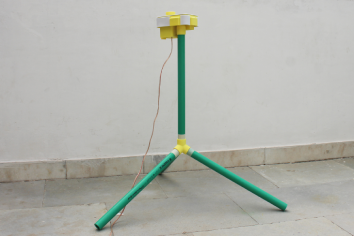}\\
    \makebox[0.3\textwidth][c]{\texttt{(m)}}\hfill
    \makebox[0.3\textwidth][c]{\texttt{(n)}}\hfill
    \makebox[0.3\textwidth][c]{\texttt{(o)}}
    \caption{Assembly steps for the GSI}
    \label{fig:assemblySteps}
\end{figure}

So far, the images are captured and stored images in the local $\mu$SDXC memory card (Part $\# 19$). However, these images must be transferred to the external storage device, i.e. the USB solid state flash drive (SSD or Part~$\# 20$) because of limited storage capacity in the memory card. This is done with the help of another script whose algorithm is depicted in the form of a flowchart as shown in the Figure~\ref{fig:SSDmanagementFC}.

With all the pre-requisites taken care of, the final assembly steps are as follows. Temporarily detach the camera from the ribbon wire that is connected to the Raspberry pi before proceeding. The alphabet(s) written next to each step within square braces correspond to the sub-figure number(s) in the Figure~\ref{fig:assemblySteps}. A more detailed exploded view of the GSI is also shown in the Figure~\ref{fig:explodedGSI} for a better visualization of the build architecture.

\begin{figure}[!ht]
    \centering
    \includegraphics[trim={0 1150 5250 0},clip,width=0.4\textwidth,keepaspectratio]{Sketches-GSIexplodedView.eps}
    \caption{Exploded view of the designed GSI}
    \label{fig:explodedGSI}
\end{figure}

\begin{enumerate}
    \item Attach the lens hood mount (Part~$\# 06$) to the main body (Part~$\# 01$) using the super glue, making sure that it is water tight. [\texttt{(a)}, \texttt{(b)}]
    \item Using $4$ $\times$ M$3$ $10mm$ bolts, attach the camera (Part~$\# 16$) to the provided studs in the main body (Part~$\# 01$). [\texttt{(c)}, \texttt{(d)}]
    \item Pick the Raspberry Pi and the completed circuit as described before. Attach the ribbon cable to the camera (Part~$\# 16$). Plug the power supply adapter (Part~$\# 15$) to the Raspberry Pi. Place the LDR sensor (Part~$\# 24$) in the slot on the Main Body and seal it with epoxy glue from both inside and outside. [\texttt{(e)}, \texttt{(f)}]
    \item Take the controller circuit, USB drive and power supply through the cavity designated on the cover (Part~$\# 02$). Close the cover onto the main body using $6$ $\times$ M$3$ $10mm$ bolts. [\texttt{(g)}, \texttt{(h)}]
    \item Connect the additional AC general purpose electrical cable to the power supply adapter and insulate the connection. Now place everything inside the accessibility chute available in the cover itself. [\texttt{(i)}]
    \item Now close the accessible chute with the lid (Part~$\# 04$) and secure it using lid retainers (Part~$\# 05$). Also attach the mount socket (Part~$\# 11$) at this point to the cover. [\texttt{(j)}]
    \item Fix the air vent filter lid or duct (Part~$\# 03$) to the cover using supper glue making sure that there is no wire underneath. [\texttt{(k)}]
    \item Cut the insulation pads from the insulation sheet (Part~$\# 13$) according to the Part~$\# 13$ design file description and paste them on the surface using a heat resistant glue. [\texttt{(l)}]
    \item Now assemble the mounting tripod stand (Part~$\# 12, 32$). [\texttt{(m)}, \texttt{(n)}]
    \item Attach the tripod stand to the mount socket to complete the GSI camera setup. Connect the AC power supply cable to any electricity port nearby and get the camera working. [\texttt{(o)}]
\end{enumerate}

\section{Operation instructions}

Once the designed GSI is completely setup, it is now ready to be installed in the field as per the requirement. While this paper provides instructions to design a cheap tripod for the GSI, it might not work in huge open areas under windy conditions. In such cases, one must redesign the stand (Part~$\# 12, 32$) and mount socket(Part~$\# 11$) according to the specific mounting conditions. An easy work around for vast open fields would be to remove the tripod completely, while keeping the mount socket intact. Then the GSI can be mounted on a straight pipe which is dug sufficiently into the ground to keep the GSI steady under strong windy conditions.

Since Raspberry Pi $4$ Model-B (Part~$\# 14$) has both gigabit Ethernet ports and WiFi adapter, it can access the internet either way. Therefore, during operations, if internet connection is available, the Raspberry Pi can be configured easily to directly backup the images onto an external server as soon as it is available. In such cases, there should not be any need to access the remote (Part~$\# 08-10$) and external SSD storage device (Part~$\# 20$) by removing the lid retainers (Part~$\# 05$) and the lid (Part~$\# 04$).

\begin{figure}[!ht]
\centering
\begin{minipage}{.5\textwidth}
  \centering
  \includegraphics[width=.8\linewidth,keepaspectratio]{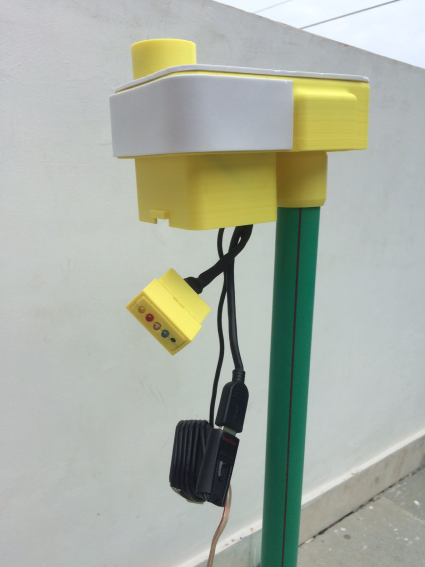}
  \caption{GSI with lid retainer and the lid removed}
  \label{fig:GSI_LidRemoved}
\end{minipage}%
\begin{minipage}{.5\textwidth}
  \centering
  \includegraphics[width=.8\linewidth,keepaspectratio]{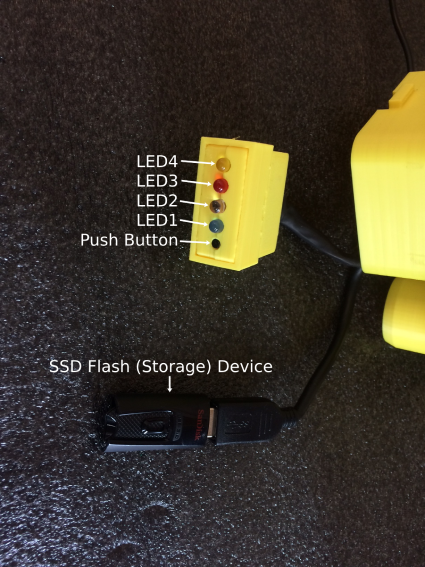}
  \caption{Remote and SSD in an isolated view}
  \label{fig:remoteDetails}
\end{minipage}
\end{figure}

However, in the case of no internet connection, the lid must be opened to access the remote and the SSD storage device and take the backup manually. Once opened, the GSI will look something like as shown in Figure~\ref{fig:GSI_LidRemoved}. To safely eject the SSD drive, and to check the status of the GSI, the remote is provided with $4$ LEDs (Part~$\# 25$) and $1$ push button (Part~$\# 29$). Referring to the terminology used in Figure~\ref{fig:SSDmanagementFC} and~\ref{fig:remoteDetails}, the use cases and meaning of each LED is as follows:
\begin{itemize}
    \item \textbf{LED1} turns ON only when Raspberry Pi detects the pressing of the push button. It will remain on until the safely ejection routine for SSD removal is completed and then will turn OFF.
    \item \textbf{LED2} is switched ON right before the safely ejection procedure starts. If SSD is not found in the system, this LED will blink $5$ times and then turn OFF. Else this LED will remain ON for $5$ seconds post completing the safely ejection routine and then it will turn OFF. One can safely remove the SSD once this LED follows the ON or ON/OFF routine and then turns OFF permanently.
    \item \textbf{LED3} turns ON before the data transfer starts happening from the local memory card of the Raspberry Pi to the external SSD flash drive. Once this procedure is completed, this LED will turn off. It is recommended not to press the push button (or initiate the SSD removal process) while this LED is turned ON.
    \item \textbf{LED4} is switched ON when the external SSD flash drive is re-detected by the Raspberry Pi. It turns OFF after $5$ seconds on its own. This is present for the cases when the SSD is reinserted into the GSI after taking the required backup manually.
\end{itemize}

Assuming that the SSD can contain up to $240GB$ of the data (\textit{i.e.} the space which is practically available in the SSD) and the internal memory card can hold up to $35GB$ of the data (i.e. excluding the space occupied by the OS and leaving space for virtual RAM, if required). This limited storage space means that the manual backup must be taken at a regular basis. Given that the images are captured at $5$ minutes interval and that each image in its maximum resolution takes up around $8MB$ of disk space, the calculations for the maximum time before which the backup must be taken are as follows:
\begin{align*}
    &\textrm{Number of images captured per day} &=& \left(\textrm{Hours/day}\right)\times\left(\frac{\textrm{Minutes/hour}}{5}\right)&\\
    \Rightarrow& \textrm{Number of images captured per day} &=& 24\times\left(\frac{60}{5}\right) &=& 288\\
    \Rightarrow& \textrm{Memory required by images from $1$ day} &=& 288\times8MB &=& 2.25GB\\
    \Rightarrow& \textrm{Number of days for which memory card can hold data} &=& \left\lfloor\frac{35}{2.25}\right\rfloor &=& 15 \textrm{ Days}\\
    \Rightarrow& \textrm{Number of days for which external SSD can hold data} &=& \left\lfloor\frac{240}{2.25}\right\rfloor &=& 106 \textrm{ Days}
\end{align*}
\vskip 0.2cm

This means that the SSD must be removed from the GSI for manual backup at least once in $3$ months (keeping a margin of around $2$ weeks) and that the SSD must be reinserted back into the GSI within the next $1$ week (keeping a margin of around $1$ week).

The lens cover is a curved glass which will automatically help slide away the residual dirt and moisture which might get deposited on it due to fluctuating weather conditions. However, it might still require proper cleaning from time to time to let the GSI capture clean images of the clouds/sky. Since the GSI does not have any automatic lens cover cleaner, one should also clean the GSI lens cover (Part~$\# 31$) manually once is $2-3$ weeks. If required, the lens hood (Part~$\# 07$) can also be removed to directly access the camera lens and the inside of the GSI lens cover to clean some residual moisture that might have seeped through air vents.

Considering that the primary use case of the designed GSI is in outdoors and even more likely, somewhere remote, there are some safety issues one must consider before installing and operating the hardware. While everything else inside the GSI is water-proof because of the casing, the electrical power supply cable is exposed to the outside environment. This cable must be safely and carefully laid down to the AC power point which should in itself be well-protected from the outside weather (like rain, snow, etc.). Additionally, the lid retainers must only be removed during clear weather conditions to access the SSD storage drive and the remote. Same goes for accessing the lens by opening the lens hood to keep the inside of the GSI away from water, dirt, snow, etc.

\section{Validation and characterization}

\begin{figure}[!ht]
    \centering
    \includegraphics[width=0.23\textwidth]{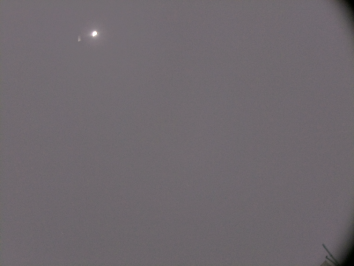}
    \includegraphics[width=0.23\textwidth]{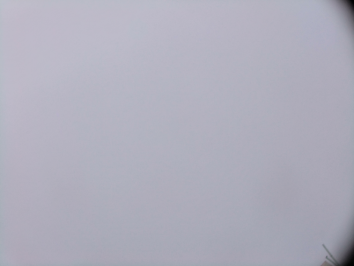}
    \includegraphics[width=0.23\textwidth]{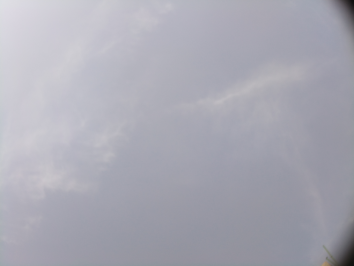}
    \includegraphics[width=0.23\textwidth]{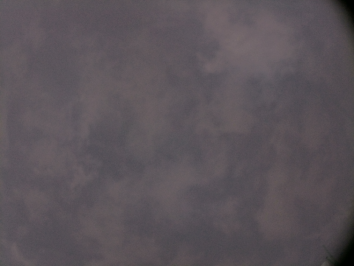}\\
    \includegraphics[width=0.23\textwidth]{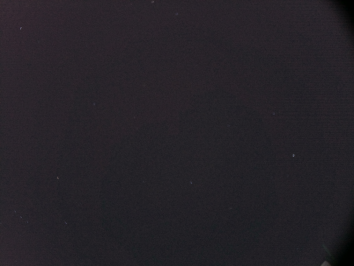}
    \includegraphics[width=0.23\textwidth]{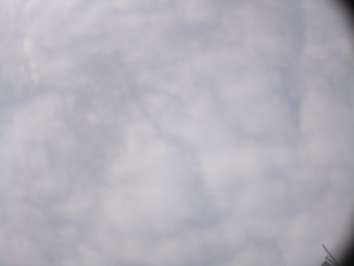}
    \includegraphics[width=0.23\textwidth]{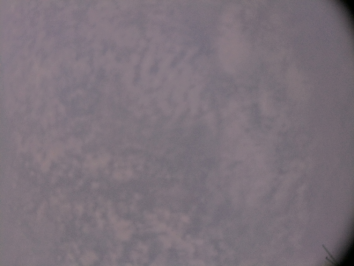}
    \includegraphics[width=0.23\textwidth]{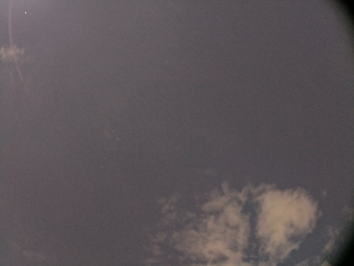}
    \caption{Sample images as captured by the designed GSI. \textit{Row-wise from left to right -} clear sky during night with moon; clear sky during day; veil clouds during day; veil clouds during night; clear sky during night without moon; thick white clouds during morning; patterned clouds during dusk; and partly clear sky during night.}
    \label{fig:GSIcaptuedImages}
\end{figure}

LAMSkyCam aims to capture sky/cloud images from both night and day time at high temporal and spatial resolution. These images are useful for various meteorological studies. Since the images from the camera are timestamped, they can be used to study cloud movements~\cite{dev2016short} and state. Figure~\ref{fig:GSIcaptuedImages} shows some of the captured sky/cloud images from different time of the day, using the designed GSI. It can be noted that while the images of the daytime are very clear (as expected), the images from the nighttime are also clean. Even the clear sky image without moon from a night successfully captures the stars as small white pixels. The fact that the images are clear at both night and day times validates the optimal performance of the camera.

The camera is already installed in Delhi, India and has been operational for nearly a month. During this time, the camera has already withstood strong winds, rain and a temperature range of $24-36^{\circ}C$. The capabilities and possible limitations of the designed GSI are summarized as follows:
\begin{itemize}
    \item Spatial resolution $=$ $4056\times3040$ pixels.
    \item Temporal resolution $=$ $5$ minutes.
    \item Operational temperature range $=$ $0-50^{\circ}C$.
    \item Automatically adjustable shutter speed based on ambient light intensity.
    \item Local (manual) backup facility for remote locations with no internet access.
    \item Specialized hardware clock to keep track of time during power outage.
    \item Miniature size and the ability to be mounted on any pole-like structure for easy installation.
\end{itemize}

We intend to scale this research as a part of our future work. With the standalone hardware instance in place, we are looking forward to use a network of such imagers to have a stereo vision~\cite{savoy2016geo} of the clouds to better estimate the heights of different layers of clouds in the atmosphere.

\section*{CRediT author statement}
\noindent
\textbf{Mayank Jain}: Conceptualization, Investigation, Software, Writing - Original Draft. 
\textbf{Vishal Singh Sengar}: Methodology, Visualization, Resources, Writing - Original Draft. 
\textbf{Isabella Gollini}: Formal Analysis, Funding Acquisition, Writing- Reviewing and Editing. 
\textbf{Michela Bertolotto}: Formal Analysis, Funding Acquisition, Writing- Reviewing and Editing. 
\textbf{Gavin McArdle}: Formal Analysis, Funding Acquisition, Writing- Reviewing and Editing. 
\textbf{Soumyabrata Dev}: Formal Analysis, Resources, Project Administration, Supervision, Funding Acquisition, Writing- Reviewing and Editing.


\section*{Acknowledgements}
\noindent
The research in this paper is funded by a research grant from Kaggle, a subsidiary of Google LLC. This research was conducted with the financial support of Science Foundation Ireland under Grant Agreement No. 13/RC/2106\_P2 at the ADAPT SFI Research Centre at University College Dublin. ADAPT, the SFI Research Centre for AI-Driven Digital Content Technology, is funded by Science Foundation Ireland through the SFI Research Centres Programme.

\noindent
\bibliographystyle{unsrt}


\noindent
\bibliographystyleURL{unsrt}

\begin{center}
---------------------------------------------------------------------------------------------------------------------------
\end{center}

\end{document}